\pdfoutput=1
\documentclass[review]{elsarticle}

\usepackage{lineno,hyperref}
\usepackage{amsmath}
\usepackage{bbold}
\modulolinenumbers[5]

\begin{document}

\begin{frontmatter}

\title{On the exact discretization of Schr\"{o}dinger equation}

\author{Chih-Lung Chou}
\address{Department of Physics, Chung Yuan Christian University, Taoyuan 320314, Taiwan}


\ead{ciwcielioong@gmail.com}

%\address[mymainaddress]{200 Zhongbei Road, Zhongli District, Taoyuan City 320314, Taiwan} 

\begin{abstract}
We show that the exact discrete analogue of Schr\"{o}odinger equation can be derived naturally from the Hamiltonian operator of a Schr\"{o}dinger field theory by using the discrete Fourier transform that transforms the operator from momentum representation into position representation. The standard central difference equation that is often used as the discretized Schr\"{o}dinger equation actually describes a different theory since it is derived from a different Hamiltonian operator. The commutator relation between the position and momentum operators in discrete space is also derived and found to be different from the conventional commutator relation in continuous space. A comparison between the two discretization formulas is made by numerically studying the transmission probability for a wave packet passing through a square potential barrier in one dimensional space. Both discretization formulas are shown to give sensible and accurate numerical results as compared to theoretical calculation, though it takes more computation time when using the exact discretization formula. The average wave number $k_0$ of the incident wave packet must satisfy $|k_0\ell| < 0.35$, where $\ell$ is the lattice spacing in position space, in order to obtain an accurate numerical result by using the standard central difference formula. 
\end{abstract}


\end{frontmatter}

\linenumbers

\section{Introduction}
In quantum mechanics the time-dependent Schr\"{o}dinger equation is a second order linear partial differential equation
\begin{eqnarray}
i\hbar \frac{\partial}{\partial t}\Psi(x,t) = -\frac{\hbar^2}{2M}\nabla^2 \Psi(x,t) + U(x)\Psi(x,t), \label{SchrodingerEq}
\end{eqnarray}
\noindent where the single-particle wavefunction $\Psi(x,t)$ is a function of position $x$ and time $t$, $U(x)$ denotes the potential energy, and $M$ is particle's mass. Though the Schrodinger equation has a specific form in terms of continuous variables $x$ and $t$, there are different discrete analogues of the equation \cite{DiscreteQMBook, discreteSE,TarasovPLA}. For example, by using the standard central difference formula for the Laplace operator 
\begin{align}
    \nabla^2 \Psi \to \sum_{a=1}^3 \frac{1}{\ell^2}\{\Psi({x}+\hat{a}\ell,t)+ \Psi({x}-\hat{a}\ell,t)-2\Psi({x},t)\},
\end{align}
\noindent a possible discretization of Eq.(\ref{SchrodingerEq})
is
\begin{align}
i\hbar \frac{\partial \Psi({x},t)}{\partial t}= -\frac{\hbar^2}{2M\ell^2}\sum_{a=1}^3 \{ \Psi({x}+\hat{a}\ell,t)+ \Psi({x}-\hat{a}\ell,t)-2\Psi({x},t)\}+U({x})\Psi({x},t). \label{discrete01}
\end{align}
\noindent Here $\Psi(x,t)$ is defined only on the discrete position ${x}=\ell(\hat{1}n_1+\hat{2}n_2+\hat{3}n_3)$ ($n_1,n_2,n_3$ are integers), $\hat{a}$ ($a=1,2,3$) denote the unit base vectors of the Cartesian coordinate system, and $\ell$ is the lattice spacing between the nearest-neighboring spatial sites. Eq.(\ref{discrete01}) cannot be thought as an exact discrete analog of Eq.(\ref{SchrodingerEq}) since it has different dispersion relation $\varepsilon(k)$ than that in the continuous Schr\"{o}dinger equation with zero-potential $U(x)=0$ \cite{discreteSE}. In other words, the theory described by (\ref{discrete01}) is different from the theory described by the exact discrete analogue of the Schr\"{o}dinger equation since both theories do have different Hamiltonian operators in discrete space. They are equal to each other only in the zero-spacing limit $\ell \to 0$. 

An exact discretization of Schrodinger equation in one-dimensional space has been derived directly from the continuous Schrodinger equation \cite{TarasovPLA}
\begin{align}
 i\hbar\frac{d\Psi({x},t)}{dt}=\frac{\hbar^2}{M\ell^2}\{\frac{\pi^2}{6}\Psi(x,t) +\mathop{\sum_{m=-\infty}^\infty}_{m \neq 0}\frac{(-1)^m}{m^2}\Psi(x-m\ell,t)\}+ U(x)\Psi(x,t). \label{exactDiscrete01}   
\end{align}
\noindent Different from the standard central difference formula, the exact discretized Schr\"{o}dinger equation has difference of integer order that is represented by infinite series, and a long-range interaction is suggested in the discretized equation. 

In this paper, we show that a natural way for the derivation of the exact discretized Schr\"{o}dinger equation is from the Hamiltonian operator of the Schr\"{o}dinger field theory. It is known that the continuous Schr\"{o}dinger equation that describes the time evolution of wavefunction can be derived from the following equation
\begin{align}
    i\hbar \frac{\partial}{\partial t}|\Psi\rangle = H|\Psi\rangle, \label{Schrodinger02} 
\end{align}
\noindent where the Hamiltonian operator $H$ plays the role of the generator of time evolution. The so-called wavefunction $\Psi(x,t)$ is the inner product between the position eigenket $|x\rangle$ and the quantum state $|\Psi\rangle$, $\Psi(x,t)=\langle x|\Psi\rangle$. Once the Hamiltonian operator is given, by taking the inner product on both sides of (\ref{Schrodinger02}), the Schr\"{o}dinger equation in (\ref{SchrodingerEq}) is obtained. Following the same line of thought, the exact discretized Schr\"{o}dinger equation could also be derived from the Hamiltonian operator of the quantum system in discrete space. For example, consider a Schr\"{o}dinger field theory with the Lagrangian density \cite{QMbook}
\begin{align}
    {\cal L} &= \frac{i\hbar}{2}\left\{\Psi^*(x,t)\frac{\partial}{\partial t}\Psi(x,t)-[\frac{\partial}{\partial t}\Psi^*(x,t)] \Psi(x,t)\right \} - \frac{\hbar^2}{2m} \nabla \Psi^*(x,t)\cdot \nabla \Psi(x,t) \nonumber \\&+U(x)\Psi^*(x,t) \Psi(x,t),\label{Lagrangian}
\end{align}
\noindent the continuous Schr\"{o}dinger equation is derived as the equation of motion in the theory. Without the self interaction term (ie., $U(x)$=0), after the second quantization of the free theory in discrete space, the free Hamiltonian operator is diagonal in the discrete momentum space 
\begin{align}
    H_0=\sum_k \varepsilon_k a_k^\dagger a_k, \label{H01}
\end{align}
\noindent where $\varepsilon_k = \hbar^2k^2/2M$ denotes the energy for a free particle with momentum $\hbar k$, and $a_k^\dagger$ and $a_k$ are the creation and the annihilation operator that satisfy the quantization relation
\begin{align}
    [a_k, a_{k'}^\dagger] = \delta_{kk'}. \label{quantization} 
\end{align}
\noindent Here the momentum vector $k$ takes only discrete values. Once the free Hamiltonian operator $H_0$ is obtained in momentum representation, it is straightforward to rewrite $H_0$ in position representation by using discrete Fourier transforms. Obviously, $H_0$ will not be diagonal in position representation due to Heisenberg's uncertainty relation, thus a spatially localized particle has a chance to hop to other locations instead of staying at the same position in the free theory. With the presence of the self interaction $U(x)$, the Hamiltonian operator becomes
\begin{align}
    H=H_0+\sum_x U(x) a_x^\dagger a_x. \label{PotentialOp}
\end{align}
\noindent Here the potential energy part of $H$ is diagonal in position representation. Once the Hamiltonian operator $H$ in position representation is obtained, the exact discrete analogue of Schr\"{o}dinger equation can be derived from (\ref{Schrodinger02}) by 
\begin{align}
     i\hbar \frac{\partial}{\partial t}\langle x|\Psi\rangle =\langle x| H|\Psi\rangle.
\end{align}

In the next section we show that the exact discretization of Schrodinger equation can be obtained by transforming $H_0$ from momentum representation into position representation. In the third section, we discuss the discrete version of the commutator relation $[\hat{X}, \hat{P}]$ between position operator $\hat{X}$ and momentum operator $\hat{P}$. Next we compare the exact discretized Schr\"{o}dinger equation with the standard central difference formula by numerically studying the problem of a wave packet passing through a potential barrier. In the last section we give our conclusion.

\section{Derivation of the exact discretized Schrodinger equation}
\begin{figure}
    \centering
    \includegraphics{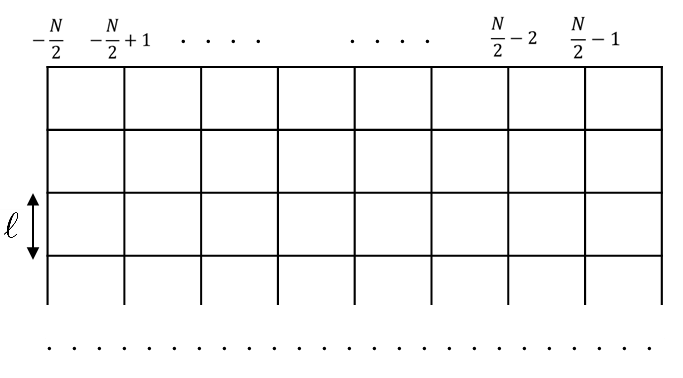}
    \caption{The three-dimensional discrete position space with a square lattice structure. The dimension in each direction is $L=N\ell$.}
    \label{fig:PositionLattice}
\end{figure}
Consider a discrete position space with a square lattice structure that is described by
\begin{align}
    {x} = \ell(\hat{1}n_1 +\hat{2}n_2+\hat{3}n_3), \quad -\frac{N}{2}\le n_1, n_2, n_3 \le\frac{N}{2}-1. \label{positionX}
\end{align}
\noindent Here ${x}$ denotes the position of the spatial lattice sites, $n_1, n_2$, and $n_3$ are integers, $\hat{a}$ ($a=1,2,3$) denote the unit base vectors of the Cartesian coordinate system, $\ell$ is the lattice spacing, and $N$ is a large even number. As shown in Fig.\ref{fig:PositionLattice}, the three-dimensional discrete position space can be viewed as a cube with side length $L=N\ell$ and total volume $L^3$. Given a free Schr\"{o}dinger field theory in the discrete space with Hamiltonian $H_0$, the operator $H_0$ is diagonal in momentum space and has the form as described in (\ref{H01}). The diagonal form of $H_0$ in momentum space is indeed a direct consequence of the translational symmetry in position space in the free theory. In the free theory, a normalized single particle state with momentum k is
\begin{align}
  |{k}\rangle = a_k^\dagger|0\rangle, \label{particleState}
\end{align}
\noindent where $|0\rangle$ denotes the vacuum state, and the momentum $k$ takes only discrete values 
\begin{align}
    {k}&=\frac{2\pi}{L}(m_1, m_2, m_3), \quad -\frac{N}{2}\le m_1,m_2,m_3 \le\frac{N}{2}-1. \label{discreteK}
\end{align}    
\noindent The integers $m_1,m_2$ and $m_3$ range from  $-N/2$ to $N/2-1$. The discrete momentum space also has a square lattice structure but with lattice spacing $2\pi/L$.

To transform Hamiltonian from momentum representation into position representation, we define the field operator $a_x$ and its hermitian conjugate $a_x^\dagger$ as the discrete Fourier transforms of $a_k$ and $a_k^\dagger$
\begin{align}
    a_x &= \frac{1}{\sqrt{N^3}}\sum_k e^{i\vec{k}\cdot \vec{x}}a_k, \label{ax}\\
    a_x^\dagger &= \frac{1}{\sqrt{N^3}}\sum_k e^{-i\vec{k}\cdot \vec{x}}a_k^\dagger. \label{axdagger}
\end{align}
\noindent Here the discrete momentum ${k}$ is defined in (\ref{discreteK}), and ${x}$ denotes the position of lattice sites in the three-dimensional discrete position space. From the quantization relation in (\ref{quantization}), the field operators $a_x$ and $a_x^\dagger$ are shown to satisfy the commutator relation
\begin{align}
    [a_x, a_{x'}^\dagger] = \delta_{xx'}. \label{QRX}
\end{align}
\noindent The commutator relation in (\ref{QRX}) thus allows us to define the position eigenkets as
\begin{align}
    |x\rangle = a_x^\dagger |0\rangle. \label{Xstate}
\end{align}
\noindent The position eigenkets are normalized states, like the momentum states $|k\rangle$, can also be single-particle states in the Schr\"{o}dinger field theory in the discrete space. From (\ref{ax}) and (\ref{axdagger}), $a_k$ and $a_k^\dagger$ can be written as linear combinations of $a_x$ and $a_x^\dagger$
\begin{align}
    a_k &= \frac{1}{\sqrt{N^3}}\sum_x e^{-i\vec{k}\cdot \vec{x}}a_x, \label{ak}\\
    a_k^\dagger &= \frac{1}{\sqrt{N^3}}\sum_x e^{i\vec{k}\cdot \vec{x}}a_x^\dagger. \label{akdagger}
\end{align}
\noindent Taking Eq.s (\ref{ak}) and (\ref{akdagger}) into $H_0$ in (\ref{H01}), we have 
\begin{align}
    H_0 = \frac{1}{N^3}\sum_{x}\sum_k \varepsilon_k a_x^\dagger a_{x} +\frac{1}{N^3}\mathop{\sum_{x,x'}}_{x\neq x'}\sum_k \varepsilon_k e^{i\vec{k}\cdot (\vec{x}-\vec{x}')}a_{x}^\dagger a_{x'}. \label{H02}
\end{align}
\noindent The form of $H_0$ in (\ref{H02}) can be simplified further if the single-particle energy $\varepsilon_k$ is known. In the Schr\"{o}dinger field theory, $\varepsilon_k = \hbar^2 {k}^2/2M$, the sum $(1/N^3)\sum_k \varepsilon_k$ in (\ref{H02}) is found to be
\begin{align}
    \frac{\hbar^2}{2MN^3}\sum_k k^2&= \frac{3\hbar^2}{2MN} \left(\frac{2\pi}{L}\right)^2\sum_{n=-N/2}^{N/2-1} n^2 
    =\frac{\hbar^2\pi^2}{2M\ell^2}(1+\frac{2}{N^2})\nonumber \\  &\longrightarrow \frac{\hbar^2 \pi^2}{2M\ell^2}, \quad \mbox{as } N\to \infty. \label{sumEk}
\end{align}
\noindent Using the fact that 
\begin{align}
    \sum_{k_a}e^{ik_a (n-n')\ell}= N\delta_{nn'}, \quad a=1,2,3.
\end{align}
\noindent where $k_a$ is the $a$-th component of wave vector $k$, $n$ and $n'$ are integers. The other sum $(1/2MN^3) \sum_k k^2 e^{i \vec{k} \cdot (\vec{x}-\vec{x}')}$ in (\ref{H02}) with nonzero spatial separation $(\vec{x}-\vec{x}')$ is found be nonzero only when $(\vec{x}-\vec{x}')$ lies in either direction $\hat{1}, \hat{2}$, or $\hat{3}$. Let assume that $\vec{x}-\vec{x}'=\hat{1} m\ell$ ($m \neq 0$), then the sum is 
\begin{align}
    &\frac{1}{2MN^3}\sum_k k^2 e^{i\vec{k}\cdot (\vec{x}-\vec{x}')} = \frac{2\pi^2}{MNL^2}\sum_{n=-N/2}^{N/2-1} n^2 (e^{i2\pi m/N})^n \nonumber \\
    &=\frac{2\pi^2}{MN^3\ell^2}\{(-1)^m\frac{N^2}{4}+ 2\sum_{n=1}^{N/2-1}n^2\cos\left({2\pi m n}/{N}\right) \} \nonumber \\&=\frac{(-1)^m\pi^2}{MN^2 \ell^2 \sin^2(m\pi/N)}, \quad m\neq 0. \label{sum02}
\end{align}
\noindent In the $N\to\infty$ limit, when $\vec{x}=\vec{x}'+\hat{a} m\ell$ ($a=1,2,3$), the sum in (\ref{sum02}) becomes 
\begin{align}
\frac{1}{2MN^3}\sum_k k^2 e^{i\vec{k}\cdot (\vec{x}-\vec{x}')} \longrightarrow \frac{(-1)^m}{M\ell^2 m^2}. \label{sum03}
\end{align}
\noindent Finally, from Eq.s (\ref{H02}, \ref{sumEk}, \ref{sum03}), the free Hamiltonian operator in the large $N$ limit in position representation is
\begin{align}
    H_0 =\frac{\hbar^2}{M\ell^2} \{\frac{\pi^2}{2} \mathbb{1}+ \sum_x \mathop{\sum_{m=-\infty}^{\infty}}_{m\neq 0} \sum_{a=1}^3 \frac{(-1)^m}{m^2} a^\dagger_{x+\hat{a}m\ell}a_x  \}. \label{H0x}
\end{align}
\noindent Here $\mathbb{1}=\sum_x a_x^\dagger a_x$ denotes the identity operator. $H_0$ is non-diagonal in position representation with the presence of the hopping interaction $a^\dagger_{x+\hat{a}m\ell}a_x$. The hopping interaction terms are responsible for the position-momentum uncertainty relation  $\Delta x \Delta p \ge \hbar/2$. With the hopping terms, a spatially localized particle can hop to other places rather than just staying at the same location. Turning on potential energy $U(x)$, the Hamiltonian operator for the Schr\"{o}dinger field theory is
\begin{align}
H=\frac{\hbar^2}{M\ell^2} \{\frac{\pi^2}{2} \mathbb{1}+ \sum_x \mathop{\sum_{m=-\infty}^{\infty}}_{m\neq 0} \sum_{a=1}^3 \frac{(-1)^m}{m^2} a^\dagger_{x+\hat{a}m\ell}a_x\}  +\sum_x U(x)a_x^\dagger a_x. \label{3DH}
\end{align}
\noindent Similarly, in two dimensional and one dimensional discrete space, the Hamiltonian operators for the corresponding Schr\"{o}dinger field theories are 
\begin{align}
    H^{2dim} &=\frac{\hbar^2}{M\ell^2} \{\frac{\pi^2}{3} \mathbb{1}+ \sum_x \mathop{\sum_{m=-\infty}^{\infty}}_{m\neq 0} \sum_{a=1}^2 \frac{(-1)^m}{m^2} a^\dagger_{x+\hat{a}m\ell}a_x  \}+\sum_x U(x)a_x^\dagger a_x, \label{2dimH0} \\
    H^{1dim} &=\frac{\hbar^2}{M\ell^2} \{\frac{\pi^2}{6} \mathbb{1}+ \sum_x \mathop{\sum_{m=-\infty}^{\infty}}_{m\neq 0} \frac{(-1)^m}{m^2} a^\dagger_{x+m\ell}a_x\}+\sum_x U(x)a_x^\dagger a_x. \label{1dimH0}
\end{align}

 The exact discrete analog of Schrodinger equation is derived as follows. Let's take the one-dimensional Schr\"{o}dinger theory as an example. Consider a single-particle quantum state $|\Psi\rangle$ in position representation 
 \begin{align}
     |\Psi\rangle = \sum_x \Psi(x,t)a_x^\dagger |0\rangle=\sum_x \Psi(x,t)|x\rangle. \label{singlePtState}
 \end{align}
 \noindent From (\ref{Schrodinger02}) and (\ref{1dimH0}), it has
 \begin{align}
&\sum_x  i\hbar\frac{\partial \Psi(x,t)}{\partial t}|x\rangle \nonumber \\
=&\frac{\hbar^2}{M\ell^2} \{\frac{\pi^2}{6} \sum_x \Psi(x,t)|x\rangle + \sum_{x} \mathop{\sum_{m=-\infty}^{\infty}}_{m\neq 0} \frac{(-1)^m}{m^2}\Psi(x,t)|x+m\ell\rangle\} +\sum_{x}U(x)\Psi(x,t)|x\rangle. \label{SchrodingerEq03}
\end{align}
\noindent Comparing the coefficient of $|x\rangle$ on both sides of the equation (\ref{SchrodingerEq03}), we get
\begin{align}
i\hbar\frac{\partial \Psi(x,t)}{\partial t} =
 \frac{\hbar^2}{M\ell^2} \{\frac{\pi^2}{6}\Psi(x,t) + \mathop{\sum_{m=-\infty}^{\infty}}_{m\neq 0} \frac{(-1)^m}{m^2}\Psi(x-m\ell,t)\}+U(x)\Psi(x,t). \label{1dimDisEq01}
 \end{align}
 \noindent The result in (\ref{1dimDisEq01}) is indeed the exact discrete analogue of Schr\"{o}dinger equation given in (\ref{discrete01}). Using the identity 
 \begin{align}
     \sum_{m=1}^\infty (-1)^m/m^2= -\frac{\pi^2}{12}, \label{12pi2}
 \end{align}
 \noindent the equation in (\ref{1dimDisEq01}) can be further written as
 \begin{align}
&i\hbar\frac{\partial \Psi(x,t)}{\partial t} \nonumber \\
=&\frac{\hbar^2}{M}\sum_{m=1}^\infty\{ \frac{(-1)^m}{m^2\ell^2}[\Psi(x+m\ell,t)+\Psi(x-m\ell,t)-2\Psi(x,t)]\}+U(x)\Psi(x,t). \label{1dimDisEq02}
\end{align}
\noindent Eq.(\ref{1dimDisEq02}) shows that the second order differential operation $\partial^2 \Psi(x,t)/\partial x^2$ can be replaced by the infinite difference 
\begin{align}
\frac{\partial^2 \Psi(x,t)}{\partial x^2} \to -2\sum_{m=1}^\infty \frac{(-1)^m}{m^2\ell^2}[\Psi(x+m\ell,t)+\Psi(x-m\ell,t)-2\Psi(x,t)], \label{infDiff}
\end{align}
\noindent in the discrete version of quantum mechanics. In two $(D=2)$ or three $(D=3)$ dimensional space, the exact discretized Schr\"{o}dinger equation is
\begin{align}
&i\hbar\frac{\partial \Psi(x,t)}{\partial t} \nonumber \\
=&\frac{\hbar^2}{M}\sum_{a=1}^D \sum_{m=1}^\infty\{ \frac{(-1)^m}{m^2\ell^2}[\Psi(x+\hat{a}m\ell,t)+\Psi(x-\hat{a}m\ell,t)-2\Psi(x,t)]\}+U(x)\Psi(x,t). \label{DdimDisEq02}
\end{align}

Now we know that Eq.(\ref{DdimDisEq02}) is an exact analogue of the continuous Schr\"{o}dinger equation since it is derived directly from the Hamiltonian operator of the theory in discrete space. The standard central difference equation given in (\ref{discrete01}) is only an approximate. In fact, Eq.(\ref{discrete01}) actually describes another quantum theory that has the Hamiltonian operator 
\begin{align}
    H=H_0+\sum_x U(x)a_x^\dagger a_x, \label{H2}
\end{align}
\noindent with
\begin{align}
    H_0&=\frac{3\hbar^2}{M\ell^2} \{\mathbb{1} -\frac{1}{6}\sum_x \sum_{b=1}^3 (a^\dagger _{x+\hat{b}\ell} a_x + a^\dagger_{x-\hat{b} \ell} a_x)\} \label{H2x}\\
    &=\frac{3\hbar^2}{M\ell^2}\sum_k [1-\frac{1}{3}\sum_{b=1}^3\cos(k_b\ell)]a_k^\dagger a_k.
    \label{H2k}
\end{align}
\noindent Obviously, the Hamiltonian operator given in (\ref{H2}, \ref{H2x}, \ref{H2k}) is different from the Hamiltonian operator given in (\ref{3DH}). The dispersion relation $\varepsilon_k$ in (\ref{H2k})
\begin{align}
    \varepsilon_k = \frac{3\hbar^2}{M\ell^2}(1-\frac{1}{3}\sum_{b=1}^3\cos(k_b\ell)), \label{dispersion}
\end{align}
\noindent is effectively the same dispersion relation given in \cite{discreteSE}. It differs from the free particle energy $\hbar^2 k^2/2M$ by less than one percent if $|k_b\ell| < 0.35$. The two Hamiltonian operators indeed describe different quantum theories in discrete space. When $|k_b\ell|<0.35$ is satisfied, the theory described by the Hamiltonian in (\ref{H2k}) could be a good approximation of the theory that is described by the Hamiltonian in (\ref{H01}). They are equal to each other only in the $\ell \to 0$ limit.

\section{Discrete analogue of canonical quantization relation}
In Schrodinger field theory in discrete space, the position and momentum operators in $\hat{b}$-direction $(b=1,2,3)$ are defined as
\begin{align}
    \hat{P}_b &= {\sum_k} \hbar k_b a_k^\dagger a_k,\quad b=1,2,3. \label{Pop}\\
    \hat{X}_b &={\sum_x} x_b a_x^\dagger a_x, \quad b=1,2,3. \label{Xop}
\end{align}
\noindent With the definition of momentum operator, a position eigenstate $|x_0\rangle$ will have a non-zero mean value of $\hat{P}_b$ 
\begin{align}
    \langle x_0|\hat{P}_b|x_0\rangle = -\frac{\hbar \pi}{L}. \label{meanP}
\end{align}
\noindent It is due to the asymmetric range of $k_b$
\begin{align}
    k_b=\frac{2\pi}{L}n_b, \quad n_b=\frac{-N}{2}, \frac{-N}{2}+1,...,\frac{N}{2}-1.
\end{align}
\noindent The non-zero mean value can be removed either by removing $k_b=-\pi/\ell$ from the allowable values of $k_b$ or by taking the limit $L\to\infty$. Similarly, a momentum state $|k_0\rangle$ also has a non-zero mean $\langle k_0|\hat{X}_b|k_0\rangle = -{\ell}/{2}$ due to the asymmetry of the range of $x_b$. It can also be removed either by removing $x_b=-L/2$ from the allowable eigenvalues of $\hat{X}_b$ or by taking the $\ell\to 0$ limit. In position representation, the momentum operator becomes
\begin{align}
\hat{P}_b =& \frac{\hbar}{N^3} \sum_{x,y}\mathop{\sum_k} k_b e^{i\vec{k}\cdot (\vec{x}-\vec{y})}a_x^\dagger a_y  \nonumber \\
=&\frac{\hbar}{N^3}\{\sum_x \mathop{\sum_k} k_b a_x^\dagger a_x + {\sum_k} \mathop{\sum_{x,y}}_{x\neq y}k_b e^{i\vec{k}\cdot(\vec{x}-\vec{y})} a_x^\dagger a_y \}  \nonumber \\
=&-\frac{\hbar \pi}{L} \{\sum_x \sum_m(-1)^m a^\dagger_{x+\hat{b}m\ell} a_x 
+i\sum_x \sum_{m\neq 0} (-1)^m \cot(\frac{m\pi}{N}) a_{x+\hat{b}m\ell}^\dagger a_x\}. \label{Pop2}
\end{align}
\noindent The first part of momentum operator in (\ref{Pop2}) 
\begin{align}
    -\frac{\hbar \pi}{L} \sum_x \sum_m(-1)^m a^\dagger_{x+\hat{b}m\ell} a_x
\end{align}
\noindent is also due to the asymmetry of the range of $k_b$. It disappears if a symmetric momentum operator is defined  \begin{align}
     \hat{P}_b &= \mathop{\sum_k}_{k_b \neq -\pi/\ell} \hbar k_b a_k^\dagger a_k,\quad b=1,2,3. \label{Popsym}
\end{align}
\noindent For spatially smooth wavefunction $\Psi(x,t)$ the first part of $\hat{P}_b$ has a negligible contribution to $\langle x|\hat{P}_b|\Psi\rangle$ since $(-1)^m\Psi(x-\hat{b}m\ell,t)$ oscillates extremely rapidly in space as $m$ increases
\begin{align}
    \langle x|\frac{\hbar \pi}{L}\sum_{x'} \sum_m (-1)^m a_{x'+\hat{b}m\ell}^\dagger a_{x'}|\Psi\rangle
   =\frac{\hbar \pi}{L} \sum_m (-1)^m\Psi(x-\hat{b}m\ell,t) \ll 1. 
\end{align}
\noindent In the $\sqrt{N}\ell \to \infty$ limit
\begin{align}
   &|\langle x|\frac{\hbar \pi}{L}\sum_{x'} \sum_m (-1)^m a_{x'+\hat{b}m\ell}^\dagger a_{x'}|\Psi\rangle| 
  \le \frac{\hbar \pi}{L}\sum_m |\Psi(x-\hat{b}m\ell,t)| \le \frac{\hbar \pi} {\sqrt{N}\ell} \longrightarrow 0. \label{1stPartP}
\end{align}
\noindent Here $\Psi(x,t)$ is assumed to be a normalized wavefunction, $\sum_x |\Psi(x,t)|^2=1$, so that $\sum_m |\Psi(x-\hat{b}m\ell,t)| \le \sqrt{N}$. Therefore, in the $\sqrt{N}\ell \to \infty$ limit for any normalized quantum state $|\Psi\rangle$, the exact discrete analogue of the differential operation becomes \cite{TarasovPLA, TarasovCommutator}
\begin{align}
    \frac{i}{\hbar}\langle x |\hat{P}_b|\Psi\rangle = \frac{\partial \Psi(x,t)}{\partial x_b} \Rightarrow \mathop{\sum_{m=-\infty}^\infty}_{m\neq 0} \frac{(-1)^m}{m \ell} \Psi(x+\hat{b}m\ell,t). \label{Pop4}
\end{align}

The commutator relation between $\hat{X}_a$ and $\hat{P}_b$ is found in a similar way. From (\ref{Pop}) and (\ref{Xop}), 
\begin{align}
    [\hat{X}_a, \hat{P}_b]&=\hbar \sum_{x}\sum_{k} x_ak_b [a_x^\dagger a_x, a_k^\dagger a_k] \nonumber \\
    &=\frac{\hbar}{N^3}\sum_{x,x'}\sum_{k} k_b(x_a-x_a') e^{i\vec{k}\cdot(\vec{x}-\vec{x}')} a_x^\dagger a_{x'}.
    \label{XP01}
\end{align}
\noindent The result in (\ref{XP01}) is not zero only when $a=b$. Without loss of generality, let's consider the case that $a=b=1$. Using
\begin{align}
    \sum_k k_1 e^{ik_1(x_1-x'_1)}&=\frac{2\pi}{L}\sum_{n=-N/2}^{N/2-1} n e^{i2\pi nm/N}=(-1)^{m+1}\frac{\pi}{\ell}(1+i\cot(\frac{\pi m}{N})),\label{sum04}
\end{align}
\noindent where $k_1=2\pi n/L$ and $x_1-x'_1=m\ell\neq 0$, the commutator $[\hat{X}_1, \hat{P}_1]$ is
\begin{align}
     [\hat{X}_1, \hat{P}_1]&=-\hbar \sum_x \sum_{m\neq 0} (-1)^m (\frac{\pi m}{N}) (1+i\cot(\frac{\pi m}{N}))a_x^\dagger a_{x-\hat{1}m\ell}. \nonumber \\
     &=i\hbar \sum_x a_x^\dagger a_x -\hbar \sum_x \sum_m (-1)^m (\frac{\pi m}{N}) (1+i\cot(\frac{\pi m}{N}))a_x^\dagger a_{x-\hat{1}m\ell}.
     \label{XP02}
\end{align}
\noindent Here the integer $m$ ranges from $-N/2 \le (x_1/\ell-n) \le N/2-1$. From the above results, we get the commutator relation
\begin{align}
     [\hat{X}_a, \hat{P}_b]=i\hbar \delta_{ab}\mathbb{1} -\hbar\delta_{ab} \sum_x \sum_m (-1)^m (\frac{\pi m}{N}) (1+i\cot(\frac{\pi m}{N}))a_x^\dagger a_{x-\hat{a}m\ell}.
     \label{XP03}
\end{align}
\noindent The commutator relation shown in (\ref{XP03}) differs from the conventional quantization relation $[\hat{X}_a, \hat{P}_b]=i\hbar\delta_{ab}\mathbb{1}$ in continuous space. Since for a Schr\"{o}dinger theory that is defined on a discrete space with finite number of lattice sites (the total number of sites is $N^3$), all momentum eigenstates $|k\rangle$ and position eigenstates $|x\rangle$ are normalized states. This means that all $|k\rangle$ and $|x\rangle$ eigenkets can be used as the single-particle quantum states in the theory if 
relativistic energy is allowed (For highly localized state $|\Psi\rangle = |x\rangle$, the particle state has energy $\langle \Psi|H|\Psi\rangle = \pi^2 \hbar^2/2M\ell^2 \gg Mc^2$). Thus, the relations that $\langle k|[\hat{X}_a, \hat{P}_b]|k\rangle=0=\langle x|[\hat{X}_a, \hat{P}_b]|x\rangle$ help to explain why the conventional commutator relation $[\hat{X}_a, \hat{P}_b]=i\hbar\delta_{ab}\mathbb{1}$ cannot be correct in discrete space.

When the commutator acts upon a quantum state $|\psi\rangle = \sum_x \Psi(x,t)|x\rangle$, it gives
\begin{align}
    \langle x|[\hat{X}_a,\hat{P}_b] |\Psi\rangle=\hbar\delta_{ab}\{i \Psi(x,t)- \sum_m (-1)^m (\frac{\pi m}{N}) (1+i\cot(\frac{\pi m}{N}))\Psi(x+\hat{a}m\ell, t) \}. \label{XP04} 
\end{align}
\noindent The infinite sum in (\ref{XP04}) has extremely small contribution when the normalized wavefunction $\Psi(x,t)$ is a localized and relatively smooth function in position space. In the situation, $(-1)^m (\frac{\pi m}{N}) (1+i\cot(\frac{\pi m}{N}))\Psi(x+\hat{a}m\ell, t)$ is not only a finite localized function but also oscillates very rapidly in space. So, in the limit that $N\gg 1$ and $\ell \to 0$, 
\begin{align}
    \sum_m (-1)^m (\frac{\pi m}{N}) (1+i\cot(\frac{\pi m}{N}))\Psi(x+\hat{a}m\ell, t) \longrightarrow 0. \label{XP05}
\end{align}
\noindent Thus in the continuous limit, for any quantum state $|\Psi\rangle$ with a localized and relatively smooth wavefunction, the conventional commutator relation holds
\begin{align}
    \langle x|[\hat{X}_a,\hat{P}_b] |\Psi\rangle=i\hbar\delta_{ab}\Psi(x,t). \label{XP06} 
\end{align}

\section{The exact discretization versus the standard central difference equation}
The problems of a Gaussian wave packet passing through one or two potential barriers have been studied either theoretically or numerically \cite{WPTunneling, TransTimeVaryB}. In the paper a comparison between the exact discrete analogue of Schr\"{o}dinger equation and the standard central difference equation is made by numerically studying the transmission probability for a Gaussian wave packet passing through a square potential barrier in one dimensional space. We assume the initial wavefunction for a particle to be a Gaussian wave packet with its center located at $x_0$
\begin{align}
    \Psi(x,t=0) = \left(\frac{1}{\sqrt{2\pi}\sigma}\right)^{1/2}\exp[-\frac{(x-x_0)^2}{4\sigma^2}+ik_0 (x-x_0)].\label{iniPsi}
\end{align}
\noindent Here $\sigma$ is the standard deviation of particle's probability distribution at time $t=0$. The initial wave packet $\Psi(x,0)$ actually consists of many plane waves with wave numbers $k$ around $k_0$ ($k_0>0$)
\begin{align}
    \Psi(x,0) = \frac{1}{\sqrt{2\pi}} \int dk \Phi(k) e^{ik(x-x_0)}, \label{iniPsik}
\end{align}
\noindent where $\Phi(k)$ is
\begin{align}
    \Phi(k)=\left(\frac{2\sigma^2}{\pi}\right)^{1/4}\exp[-\sigma^2(k-k_0)^2]. \label{iniPhi}
\end{align}
\noindent Since $\Phi(k)$ is symmetric about $k_0$, the average energy for the wave packet is thus $E_0=\hbar^2k_0^2/2M$. The wavefunction at a later time is obtained by solving Schr\"{o}dinger equation. Using the exact discretized Schr\"{o}dinger equation in (\ref{DdimDisEq02}), the wave function $\Psi(x,t+\Delta t)$ at a later time $t+\Delta t$ can be obtained from the same wavefunction at the previous time $t$
\begin{align}
    \Psi(x,t+\Delta t) &= \Psi(x,t) -i\Delta \tau \{ \sum_{m>0}\frac{(-1)^m}{m^2}[\Psi(x+m\ell,t)+\Psi(x-m\ell,t) \nonumber \\
    &-2\Psi(x,t)] +\eta(x) \Psi(x,t)  \}+O((\Delta \tau)^2), \label{Num01}
\end{align}
\noindent where $\Delta \tau \equiv \hbar \Delta t/(M\ell^2)$ denotes the time step parameter in numerical calculation, and $\eta(x)$ is defined as
\begin{align}
    \eta(x)=\frac{M\ell^2 U(x)}{\hbar^2}=\frac{(k_0\ell)^2 U(x)}{2E_0}. \label{eta}
\end{align}
\noindent As long as $\Delta \tau$ is small enough the numerical calculation based on (\ref{Num01}) should give an accurate and sensible result, and does not depend on the lattice spacing $\ell$ directly. On the other hand, if the standard central difference equation is used in numerical calculation, then $\Psi(x,t+\Delta t)$ is merely determined by the wavefunction at the previous time  $t$ at site $x$ and its nearest neighbors $x\pm \ell$
\begin{align}
    \Psi(x,t+\Delta t) &= \Psi(x,t) -i\Delta \tau \{\Psi(x,t)-\frac{1}{2}[\Psi(x+\ell,t)+\Psi(x-\ell,t)] \nonumber \\
    & +\eta(x) \Psi(x,t)  \}+O((\Delta \tau)^2). \label{Num02}
\end{align}
\noindent Numerical calculation based on (\ref{Num02}) needs not only the enough small value of $\Delta \tau$ but also the condition $|k_0\ell|<0.35$. Unlike the use of the exact discretization in (\ref{Num01}), a suitable choice of the spatial lattice spacing $\ell$ is important in getting a reasonable numerical result. The condition $|k_0\ell|<0.35$ guarantees that the theory described by the standard central difference equation is close to the conventional Schr\"{o}dinger field theory. Actually, the discrepancy in free particle's energy between the two theories is less than one percent if $|k_0\ell|<0.35$ is satisfied. 
From the aspect of numerical calculation, the most important difference between the two discretization methods is the non-local nature in the exact discretization method. In (\ref{Num01}), $\Psi(x,t+\Delta t)$ is determined by the same wavefunction at all spatial lattice sites in the previous time moment $t$. It is very different from the standard central discretization in which $\Psi(x,t+\Delta t)$ is only determined by the wavefunction at the same and the nearest neighboring sites in the previous time $t$. Thus a numerical calculation based on the exact discretization formula usually takes a much longer execution time than that based on the standard central difference equation.

\begin{figure}
    \centering
    \includegraphics{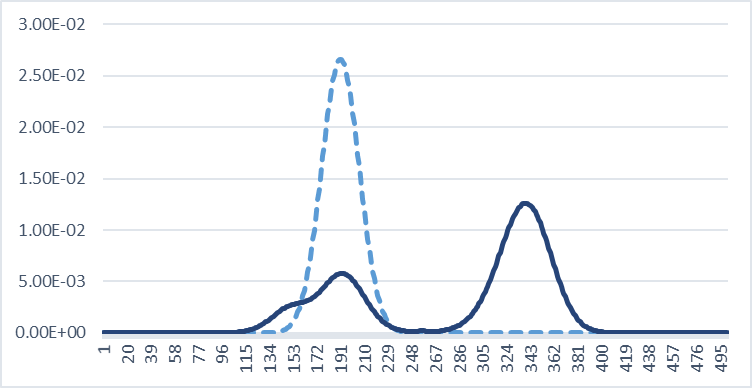}
    \caption{Quantum scattering of a wave packet through a potential barrier of height $U$ and width $W=10 \ell$, with $\ell=1$. The bell-shaped incident wave packet (probability distribution: dashed line) has the average energy $E_0$ with $E_0/U=\pi^2/8.0$, $k_0=\pi/6$, and $\sigma=15\ell$.  The reflected and transmitted wave (probability distribution) is shown as the solid line in the figure. The potential locates between the lattice site 251 and 261.  }
    \label{fig:BarrierTrans}
\end{figure}

Fig.\ref{fig:BarrierTrans} shows the reflected and transmitted wave of an incident wave packet passing through a potential barrier that is located at the center of space with height $U$ and barrier width $W=10\ell$. The space is one-dimensional with $N=500$ lattice sites and the lattice spacing is set as $\ell=1$. The average energy $E_0$ for the incident particle is chosen as $E_0/U=\pi^2/8.0$ with $k_0=\pi/6$, and the standard deviation $\sigma$ of the probability distribution of the initial wave packet is $\sigma=15 \ell$.  By choosing the time step $\Delta \tau=0.001$, the numerical calculation based on (\ref{Num01}) gives the transmission probability
\begin{align}
    \mbox{P}(transmission) = 0.654,
\end{align}
\noindent a value that is close to $0.632$ by theoretical calculation
\begin{align}
    \mbox{P}_{th}(transmission) = \int dk \frac{4\varepsilon_k(\varepsilon_k-U)|\Phi(k)|^2}{4\varepsilon_k(\varepsilon_k-U)+U^2\sin^2(\alpha(k) W)},
\end{align}
\noindent where $W$ denotes the width of the square potential, and
\begin{align}
    \alpha(k)=\sqrt{\frac{2M(\varepsilon_k-U)}{\hbar^2}}
\end{align}
could be either real or imaginary depending on whether $(\varepsilon_k-U)$ is positive or negative. The numerical calculation which is based on the standard central difference equation gives $0.603$ for the transmission probability, about $5$ percents lower than the theoretical value. The accuracy can be enhanced further to give $0.633$ for the transmission probability if $\ell=1/3$ is chosen and the space has $1500$ lattice sites. The value of $k_0\ell$ is reduced from $\pi/6>0.35$ for $\ell=1$ to $\pi/18<0.35$ for $\ell=1/3$. Smaller value of $k_0\ell$ usually leads to the numerical result with higher accuracy as expected.

Other than the aspect of numerical calculation, the exact discretized Schr\"{o}dinger equation does present a non-local transport behavior of particles. In the Schr\"{o}dinger field theory, a spatially localized quantum particle at location $x$ can jump to any location $x+m\ell$ ($m\neq 0$) with a probability in a short time $\Delta t$
\begin{align}
    \mbox{Prob}(x \to x+m\ell)=\left(\frac{\hbar \Delta t}{M m^2\ell^2}\right)^2, \label{QJump}
\end{align}
\noindent not just jump to its nearest neighboring sites. With the probability, the expected value of $m^2\ell^2$ is
\begin{align}
    \langle m^2\ell^2 \rangle = \left( \frac{\hbar \pi \Delta t}{\sqrt{3}M\ell}\right)^2. \label{ml2}
\end{align}
\noindent It implies that, in a short time interval $\Delta t$, the standard deviation of the jumping distance for the highly localized particle is $\hbar \pi \Delta t/(\sqrt{3}M\ell)$. Thus the standard deviation of the momentum $P$ for the localized particle is $\Delta P = \hbar\pi/(\sqrt{3}\ell)$, and the uncertainty relation holds
\begin{align}
    (\Delta P)\ell =\frac{\hbar \pi}{\sqrt{3}}>\Delta P \Delta x \ge \frac{\hbar}{2}. \label{DPDx}
\end{align}

\section{Conclusion}
In this paper we show that the exact discrete analogue of Schr\"{o}dinger equation can be derived naturally from the Hamiltonian operator that is given in momentum representation after the second quantization of the Schr\"{o}dinger field theory. By defining the field operators $a_x$ and $a_x^\dagger$ to be the discrete Fourier transforms of $a_k$ and $a_k^\dagger$, the Hamiltonian operators can be transformed into that in position representation. The exact discretized Schr\"{o}dinger equation is then easily derived from time evolution of quantum states. The position and momentum operators are also constructed in position representation in the paper. The commutator relation between the two operators in discrete space is also derived and found to be different from the conventional commutator relation in continuous space. This results from the fact that both momentum and position eigenkets can be the single-particle quantum states for the Schr\"{o}dinger field theory in discrete space. In the continuous limit ($N\to \infty, \ell\to 0$), the commutator relation in discrete space is shown to go back to the conventional one in continuous space. Though in the paper we assume that the creation and annihilation operator satisfy bosonic quantization relation $[a_k, a_{k'}^\dagger]=\delta_{kk'}$, the results in the previous sections including the exact discrete analogue of Schr\"{o}dinger equation, and the commutation relation between position and momentum operator, are remained the same with fermionic quantization relation $\{a_k, a_{k'}^\dagger\} = \delta_{kk'}$. 

A comparison between the exact discrete analogue of Schrodinger equation and the standard central difference equation is made by numerically studying the transmission probability for a particle passing through a square potential barrier in one dimensional space. In the quantum scattering problem, both discretization schemes give sensible and accurate results as compared to theoretical calculation. Usually it will take more computation time when using exact discretization formula in numerical calculation since it needs one to sum up contributions from the wavefunctions at all spatial sites. On the other hand, the condition $|k_0\ell|<0.35$ must be satisfied if we want to have an accurate numerical result by using the standard central difference equation. Sometimes it means that a smaller lattice spacing $\ell$ and thus a larger number of spatial lattice sites must be chosen in numerical calculation. 

Conceptually, the exact discretized Schr\"{o}dinger equation is more quantum-like than the standard central discretization since it allows particles to jump to remote sites via the hopping interaction terms $a_{x+m\ell}^\dagger a_x$ in the Hamiltonian operator. The hopping interactions origin from the free Hamiltonian operator and play important roles in particle transport. The theory that leads to the standard central discretization simply does not have such hopping terms except for the jumps to nearest neighboring sites $a_{x\pm\ell}^\dagger a_x$.

\end{document}